\documentclass[twocolumn]{aastex631}

\usepackage{xcolor}
\usepackage{amsmath}
\usepackage{float}
\usepackage{hyperref}

\begin{document}

\title{Identification and analysis of galactic bars in DESI Legacy Imaging Surveys}

\author{Wenwen Wang}
\affiliation{National Astronomical Observatories, Chinese Academy of Sciences, Beijing 100101, People’s Republic of China}
\affiliation{College of Astronomy and Space Sciences, University of Chinese Academy of Sciences, Beijing 100049, People’s Republic of China}

\author[0000-0002-4135-0977]{Zhimin Zhou}
\affil{National Astronomical Observatories, Chinese Academy of Sciences, Beijing 100101, People’s Republic of China}
\correspondingauthor{Zhimin Zhou}
\email{zmzhou@nao.cas.cn}





\begin{abstract}
We present a comprehensive study of bar structures in the local Universe using data from the DESI Legacy Imaging Surveys. Through isophotal analysis of 232,142 galaxies, we identify bars and classify them into strong and weak categories based on normalized bar length, using a threshold of 0.4. We find a total bar fraction of 42.9\%, rising to 62.0\% in disk galaxies, with strong barred galaxies accounting for 30.0\%. For barred galaxies in our sample, deprojected bar lengths are measured both in absolute terms and normalized by galaxy size. Most bars are found to have absolute lengths of 3–7 kpc, and normalized bar lengths concentrated around a median value of 0.4. Bar ellipticity mainly ranges from 0.2 to 0.6, with a median value of 0.3.
Our analysis reveals a bimodal distribution of bar fractions with respect to galaxy color, with weak bars in our classification being more prevalent in intermediate-color systems. With respect to stellar mass, strong bars also present a bimodal distribution, while weak bars are distributed uniformly. Normalized bar length remains relatively stable across stellar masses, while absolute bar length positively correlates with stellar mass.
Cross-validation with visual classifications from GZD catalog confirms a bar identification accuracy of 93\%. These results validate our automated method for bar identification and measurement, demonstrating its reliability. Our findings underscore the importance of bars in galaxy evolution and highlight the potential of upcoming wide-field surveys to deepen our understanding of barred galaxies.

\end{abstract}



\section{Introduction} \label{sec:intro}

Bars are one of the most prevalent structures in the nearby Universe, with observations indicating that roughly two-thirds of spiral galaxies possess them \citep{2002MNRAS.336.1281W, 2008ApJ...675.1141S, 2011MNRAS.411.2026M, 2018MNRAS.474.5372E}. Of these, about one-third display prominent or strong bars \citep{2004ApJ...615L.105J, 2009ApJ...698.1639M, 2012ApJ...745..125L}. Typically originating from the galactic center, bars play a critical role in redistributing gas and energy throughout the galaxy’s disk, significantly influencing its dynamics and evolution \citep{2019MNRAS.484.5192C}. The formation of bars is generally attributed to both internal mechanisms, such as intrinsic disk instabilities \citep{2002ApJ...569L..83A, 2017ApJ...839...24K, 2018MNRAS.473.2608Z}, and external influences, including intergalactic interactions, disturbances, and tidal effects  \citep{2004MNRAS.347..220B, 2014MNRAS.445.1339L}.

Theoretically, the impact of bars on their host galaxies is primarily attributed to the non-axisymmetric component of the gravitational potential \citep{2008ApJ...675.1141S, 2013MNRAS.429.1949A, 2015MNRAS.447.4018G}. As key drivers of galactic evolution, bars not only shape the morphology of galaxies but also are correlated with various properties, including size, mass, stellar age, metallicity, and gas content \citep{1995A&A...301..649F, 2011MNRAS.415.3627H, 2011MNRAS.416.2182E, 2012MNRAS.424.2180M, 2013ApJ...779..162C}. Furthermore, the bar fraction exhibits notable evolution with redshift. Early studies have shown that the fraction of barred galaxies declines as redshift increases up to around  z $\approx$ 1 , after which it stabilizes with minimal changes \citep{2008ApJ...675.1141S, 2014MNRAS.438.2882M, 2014MNRAS.445.3466S}. However, recent observations from the James Webb Space Telescope (JWST) have identified an increasing number of high-redshift barred galaxies \citep{2022ApJ...939L...7C, 2023ApJ...955...94F, 2023ApJ...945L..10G, 2023ApJ...948L..13J, 2023Natur.623..499C, 2024MNRAS.530.1984L}, indicating that new findings and insights into bar formation and evolution may soon emerge.

However, significant uncertainties still persist in our understanding of galactic bars, even within the local Universe. These research discrepancies can be attributed to two main factors: the inherently complex mechanisms governing the formation and evolution of bars, and the variability in sample selection, data quality, and methodological approaches used across different studies.

\begin{figure*}[htp!]
\centering
\includegraphics[width=0.7\linewidth]{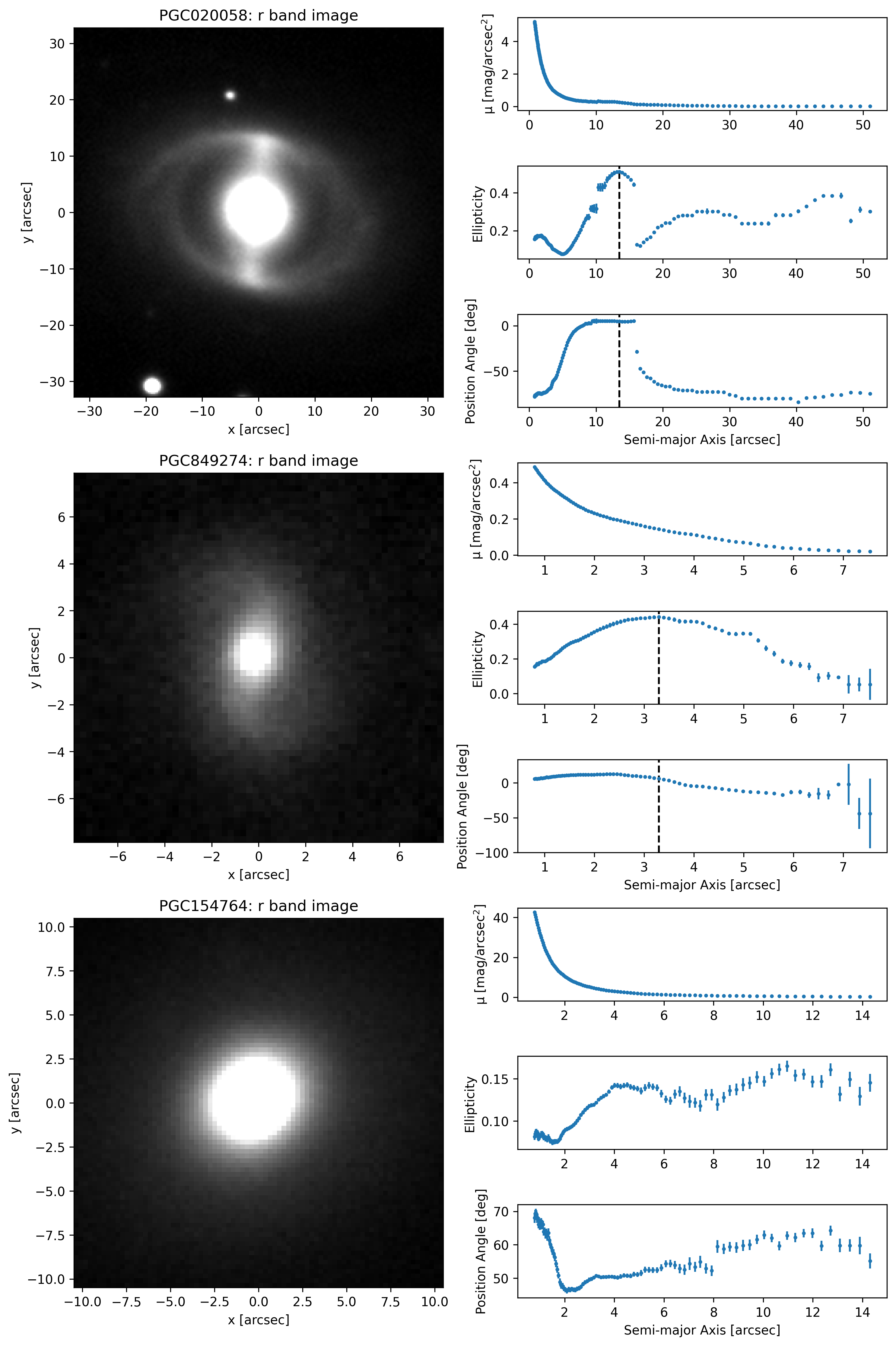}
\caption{Isophotal fitting for three example galaxies from SGA-2020: a strong bar (PGC020058), a weak bar (PGC849274), and no bar (PGC154764). The left panel for each galaxy displays the r-band image, with the size adjusted relative to the source. Both axes are scaled in arcsecond. The right panel shows radial profiles of surface brightness ($\mu$), ellipticity, and position angle as functions of the semi-major axis. The black dashed lines indicate the bar ends. Position angle is measured counterclockwise from the +y axis. See Section 3.1 for further details.
\label{fig:example}}
\end{figure*}

The main methods for identifying galactic bars currently include visual inspection \citep{2015MNRAS.449..820W}, machine learning \citep{2020MNRAS.493.4209C}, Fourier analysis \citep{2019ApJ...872...97L}, and isophote fitting \citep{2011ApJS..197...22L}. Visual inspection, as the name implies, involves manually examining the morphological features of galaxies for classification. It is one of the earliest and most traditional techniques, relying entirely on human judgment. Prominent galaxy catalogs such as RC3 \citep{1991S&T....82Q.621D}, Ann15 \citep{2015ApJS..217...27A} and Nair10 \citep{2010ApJS..186..427N} have been compiled using this approach. Among visual inspection projects, Galaxy Zoo stands out as the most notable example \citep{2008MNRAS.389.1179L}. This large-scale citizen science initiative engages volunteers to classify galaxy morphologies, using their collective input to estimate the likelihood of specific features, such as bars \citep{2011MNRAS.411.2026M,2013MNRAS.435.2835W,2021MNRAS.507.4389G}. Remarkably, the classifications from Galaxy Zoo have shown high consistency with those made by professional astronomers \citep{2014ARA&A..52..291C}. While visual inspection has historical significance and notable strengths,  the results benefit from further validation to mitigate potential subjective bias.

Another traditional approach to morphological classification involves fitting and analyzing the light profiles of galaxies using their 2D images. Fourier analysis is a prominent example of this method. It employs Fourier decomposition to study the azimuthal brightness distribution along concentric ellipses in the galaxy’s projected image \citep{1990ApJ...357...71O,2000A&A...361..841A}. By applying a two-dimensional Fast Fourier Transform (FFT), the amplitudes of different Fourier moments can be quantified, allowing the estimation of bar strength. Additionally, the variation in the second-order Fourier moment is used to determine the bar length.

A closely related technique is isophote analysis, which is primarily used to determine the surface brightness profiles of galaxies. This method involves fitting a series of concentric ellipses to the 2D images of galaxies and identifying bars based on the analysis of geometric parameters such as ellipticity and position angle \citep{1999BaltA...8..535M}.

With the rapid development of technology, automated classification methods like machine learning and deep learning have become increasingly prominent. These approaches enable computers to autonomously learn data patterns, make predictions, and make informed decisions. In astronomy, machine learning has proven to be a powerful in the analysis of large-scale datasets. For instance, \citet{2022MNRAS.509.3966W} utilized deep learning models trained on Galaxy Zoo DECaLS volunteer votes, producing comprehensive morphological catalogs for over 314,000 galaxies. As the successor of this study, Galaxy Zoo DESI (hereafter GZD) provide automated classifications for 8.7 million galaxies in the The Dark Energy Spectroscopic Instrument (DESI) Legacy Imaging Surveys and present detailed morphology measurements for them \citep{2023MNRAS.526.4768W}.

Different research methods can yield varying results when analyzing galactic bars. For example, studies using visual inspection or Fourier analysis have generally reported a higher bar fraction in early-type spiral galaxies \citep{2008ApJ...675.1141S, 2009ApJ...692L..34L, 2013ApJ...779..162C}. In contrast, results from isophote fitting often show lower bar fractions for the same galaxy types \citep{2008ApJ...675.1194B, 2009A&A...495..491A}. In a recent study, \cite{2024A&A...683A.100G} applied multiple techniques to measure the length and strength of simulated bars, finding significant discrepancies between the methods. Specifically, bar lengths estimated via Fourier analysis were notably shorter compared to those derived from isophote fitting. Moreover, \cite{2024MNRAS.52711777M} investigated bars across multiple wavelengths, from ultraviolet to infrared, and discovered that bar properties change significantly with wavelength. In some bands, bars were not even detectable, underscoring the impact of wavelength choice on bar identification.

In this paper, we conduct a comprehensive investigation of bar structures in the local Universe, focusing on their identification and detailed analysis. Using data from the DESI Legacy Imaging Surveys, we perform isophotal analysis on a large sample of nearby galaxies. We calculate the bar fraction, characterize the properties of bars, and examine their correlations with the properties of their host galaxies. To ensure the robustness and reliability of our results, we also apply cross-validation techniques to assess the robustness of our results.

This paper is structured as follows: In Section \ref{sec:data}, we describe the data and sample selection criteria used in our study. Section \ref{sec:methods} outlines the methods and procedures employed for bar identification. Our results on bar classification and properties are presented in Section \ref{sec:results}, followed by a discussion of their implications in Section \ref{sec:discussion}. Finally, in Section \ref{sec:summary}, we summarize our findings and conclude the paper.

\begin{table*}
\renewcommand{\arraystretch}{1.2}
\begin{rotatetable*}
\begin{center}
\caption{Properties of Bars \label{tab:d}}
\begin{tabular}{ccccccccc}
\hline
SGA\_ID & GALAXY & RA & DEC & G\_MAG & R\_MAG & Z\_MAG & Bar\_Length & Bar\_Ellipticity\\
 & & ($\rm degree$) & ($\rm degree$) & ($\rm magnitude$) & ($\rm magnitude$) & ($\rm magnitude$) & & \\
\hline
2& PGC1283207 & 228.38 & 5.42 & 16.71 & 15.86 & 15.24 & 0.61 & 0.19\\
207& SDSSJ113046.32+673514.6 & 172.69 & 67.59 & 18.01 & 17.36 & 16.89 & 0.26 & 0.13\\
19161& PGC1317770 & 165.13 & 7.27 & 16.45 & 15.61 & 15.00 & 0.43 & 0.30\\
175294& PGC638448 & 19.43 & —36.30 & 16.90 & 16.04 & 15.42 & 0.38 & 0.58\\
242137& PGC1825224 & 356.32 & 28.06 & 15.39 & 14.48 & 13.80 & 0.47 & 0.34\\
306395& PGC1150606 & 37.06 & —0.15 & 17.24 & 16.85 & 16.61 & 0.30 & 0.37\\
424178& NGC0977 & 38.26 & -10.76 & 13.69 & 12.96 & 12.40 & 0.26 & 0.45\\
475380& PGC1486113 & 203.59 & 15.49 & 16.57 & 15.83 & 15.30 & 0.44 & 0.27\\
524838& PGC520194 & 30.96 & —46.00 & 16.43 & 15.54 & 14.91 & 0.64 & 0.14\\
668985& PGC041704 & 188.35 & 46.63 & 15.22 & 14.69 & 14.44 & 0.30 & 0.36\\
790607& PGC1048787 & 187.61 & -5.14 & 15.92 & 15.06 & 14.44 & 0.68 & 0.64\\
835418& PGC032375 & 162.32 & 0.33 & 15.01 & 14.28 & 13.74 & 0.39 & 0.53\\
952759& IC1076 & 223.75 & 18.04 & 14.06 & 13.51 & 13.13 & 0.28 & 0.37\\
1066793& PGC1459038 & 343.50 & 14.47 & 16.71 & 15.80 & 15.17 & 0.49 & 0.26\\
1266484& PGC073895 & 19.00 & 0.04 & 16.70 & 15.98 & 15.44 & 0.28 & 0.45\\
1266548& PGC1243761 & 211.99 & 3.04 & 16.57 & 15.73 & 15.11 & 0.25 & 0.35\\
1433399& 2MASXJ11371725+3301457 & 174.32 & 33.03 & 16.79 & 16.10 & 15.63 & 0.34 & 0.30\\
1433576& UGC01705 & 33.30 & 9.51 & 15.88 & 15.27 & 14.80 & 0.24 & 0.82\\
4003034& IC2900 & 172.87 & 13.17 & 16.75 & 16.13 & 15.67 & 0.36 & 0.69\\
5000992& DR8-2281m042-158 & 228.19 & —4.37 & 17.78 & 16.89 & 16.18 & 0.54 & 0.49\\
\hline
\end{tabular}
\end{center}
\tablecomments{The bar properties for 20 randomly selected galaxies. Column 1: numerical index in SGA-2020. Column 2: galaxy name. Columns 3–4: right ascension and declination. Columns 5-7: cumulative brightness measured within semi-major axis length at the 25 $mag\,arcsec^{-2}$ surface brightness isophote in $\it g$, $\it r$, $\it z$ bands (-1 if not measured). Columns 8: deprojected bar length in the r-band normalized by half of D25\_LEDA which is the approximate major-axis diameter at the 25 $mag\,arcsec^{-2}$ surface brightness isophote. Columns 9: deprojected bar ellipticity. The units are noted in the table. The full version of this table is available at ScienceDB: \dataset[10.57760/sciencedb.19973]{\doi{10.57760/sciencedb.19973}}.}
\end{rotatetable*}
\end{table*}

\section{Data \& Sample} \label{sec:data}
\subsection{Sample Selection}
DESI Legacy Imaging Surveys \citep{2019AJ....157..168D} is an extensive observational program combining three major sky surveys: the Mayall z-band Legacy Survey (MzLS), the Dark Energy Camera Legacy Survey (DECaLS), and the Beijing-Arizona Sky Survey (BASS). This project covers approximately 14,000 square degrees of the extragalactic sky observable from the Northern Hemisphere, imaging three optical bands ($\it g$, $\it r$, $\it z$) and supplementing with four mid-infrared bands (3.4, 4.6, 12, and 22 $\mu$m) from the NEOWISE mission. The latest data release (DR10) further extends the coverage to over 20,000 square degrees by incorporating additional DECam data from NOIRLab in optical bands ($\it g$, $\it r$, $\it i$, $\it z$).

Using data from the DESI Legacy Imaging Surveys, \citet{2023ApJS..269....3M} employed a probabilistic inference technique to estimate the shapes and brightness profiles of astronomical sources, leading to the development of the Siena Galaxy Atlas 2020 (SGA-2020). This extensive multi-wavelength optical and infrared atlas features a catalog of 383,620 nearby galaxies, achieving over 95\% completeness for galaxies larger than $R$(26) $\approx$ 25$''$ and with magnitudes $r < 18$. The atlas provides a wealth of detailed information, including precise coordinates, high-quality multi-band mosaics, surface brightness profiles, and comprehensive photometric data, making it an invaluable resource for studies of galaxy morphology and structural analysis.

In this study, we utilized galaxies from the Siena Galaxy Atlas 2020 (SGA-2020) as our primary sample. The images are presented in "custom brick" based on the estimated center and diameter of the galaxy group from DESI imaging. Specifically, the mosaic diameter for each galaxy is 2–3 times the size of the galaxy group. The projections for the $g$, $r$, $z$ filters are identical, with a pixel scale of 0.262$''$/pixel. To enhance the accuracy of structural identification and analysis, we focused on galaxies to have a moderate inclination ($i \lesssim 60^{\circ}$). Specifically, we selected galaxies with an axial ratio $b/a > 0.5$, where $\it b$ and $\it a$ represent the semi-minor and semi-major axes, respectively. This criterion, which is widely adopted in previous research \citep{2007ApJ...657..790M,2008ApJ...675.1194B,2011ApJS..197...22L}, ensures an ellipticity  $\textit{e}=1-b/a$ of less than 0.5, minimizing the effects of projection and inclination. By applying this selection filter, our final sample consists of 232,142 galaxies, covering a redshift range of $z \approx$ 0 to 0.4.

\subsection{Ancillary Datasets}
\subsubsection{Cross-checking} \label{sec:2.2.1}
We also utilize the GZD catalog to validate our results. As noted earlier, GZD provides detailed morphological catalogs for galaxies from the DESI Legacy Imaging Surveys, encompassing automated morphology predictions using deep learning models. The data were all collected from DESI project which means the overlap ensures a high degree of compatibility between the GZD catalog and our sample.

GZD catalog includes results encompassing automated measurements for 8.67 million galaxies. Their models are trained on newly collected votes for DESI-LS DR8 images as well as historical votes from GZ DECaLS. The automated catalog also provides predicted probabilities for various morphological features, such as disks, the presence of bars, spirals arms and ongoing mergers, mirroring the volunteer responses.

We constructed a comparison sample based on GZD. We cross-matched our selected galaxies with the GZD catalog using positional coordinates, applying a search radius of 1 arcsecond to minimize positional discrepancies. This initial matching resulted in a sample of 202,322 galaxies.

\subsubsection{Galaxy Properties}
We enhanced the SGA-2020 catalog by incorporating high-quality, homogeneous catalogs of galaxy physical properties from existing literature. Specifically, we obtained stellar masses and photometric redshifts from the catalog provided by \citet[hereafter Z19]{2019ApJS..242....8Z}, which includes data for 170 million galaxies from the DESI Legacy Imaging Surveys. The photometric redshifts (photo-$\it z$) in this catalog were estimated using a training-based method that establishes a relationship between redshift and the photometric spectral energy distribution (SED) in local color space. Z19 achieved redshift estimates for approximately 2.2 million galaxies with an accuracy of around 0.017, indicating high reliability. Additionally, by fitting the SEDs of galaxies, they derived stellar masses and other physical properties based on the estimated photometric redshifts.

\section{Methods} \label{sec:methods}
\subsection{Isophotal Analysis}
We employ isophotal analysis on 2D images to identify bars in galaxies. Given that the isophotes of most galaxies, particularly early-type systems, closely resemble ellipses, we use a set of ellipses to fit the isophotes based on the iterative method described by \citet{1987MNRAS.226..747J}. In this process, ellipses are sampled along the semi-major axis of the galaxy, with either fixed or variable step sizes, and their radii increase progressively until they reach a predefined outer boundary. This approach effectively differentiates various galaxy types and structures by capturing variations in their surface brightness profiles. Additionally, the isophotal analysis provides comprehensive data on surface brightness, ellipticity, position angle, and central coordinates. These parameters fulfill most research needs related to galaxy morphology and are particularly useful for bar identification.

In this study, we utilize the \textit{ellipse} task in \textit{IRAF} to perform isophote fitting on our sample galaxies. Initial estimates for multiple geometric parameters, such as the coordinates of the fitting ellipse center, ellipticity (\textit{e}), and position angle (PA) of the galaxy, are obtained from the photometric data of SGA-2020. These initial values are then fine-tuned in the fitting procedure, including adjustments for step size and other necessary parameters. The \textit{ellipse} task automatically generates the radial fitting results along with associated error estimates.

The key parameters for identifying galactic bars include ellipticity, position angle, and their respective variations. \citet{1995A&AS..111..115W} demonstrated that bars are typically characterized by an increase in ellipticity and a nearly constant PA. We adopted the refined criteria initially proposed by \citet{2004ApJ...615L.105J}, which have been widely utilized in subsequent studies \citep{2005MNRAS.364..283E, 2007ApJ...657..790M, 2008MNRAS.384..420G, 2009A&A...495..491A, 2011ApJS..197...22L, 2023ApJ...945L..10G}. According to these criteria, we define a bar as follows: First, within the bar-dominated region, ellipticity (\textit{e}) should gradually increase, reaching a peak value (\textit{e}$_{\rm max}$) greater than 0.25, while the PA remains relatively stable, with the variation $\rm PA_{var}$ no more than $20^{\circ}$. Second, at the end of the bar, ellipticity must decrease by at least 0.1 from the peak value (denoted as $e_{\rm end\_change}$), and the PA should show a change of more than $10^{\circ}$ (denoted as $\rm PA_{end\_change}$. The bar length, $R_{b}^0$, is defined as the semi-major axis length corresponding to \textit{e}$_{\rm max}$, representing the observed bar ellipticity.

Taking advantage of the basic criteria, we have developed a tool to automatically identify the bars in galaxies \footnote{The code is available at \url{ https://github.com/zhmzhou/identifybars}}.
In this tool, we incorporate additional flexible criteria to account for complicating effects caused by disk artifacts, such as spiral arms or clumps, which can potentially contaminate or mimic bar signatures. For example, the change rates of \textit{e} and PA along the radius within the potential bar regions ($e_{\rm var}$ and $\rm PA_{var}$) are calculated and incorporated into the analysis. Based on these estimated parameters, we establish a series of criteria, including both classic methods e.g., ($e_{\rm max} > 0.25$, ${e_{\rm end\_change}} \le 0.1$, $\rm {PA_{end\_change}} \geq 10^\circ $, $\rm PA_{var} \le 20^\circ $) and empirical approaches (e.g., $e_{\rm var} > 0.01\ \rm pixel^{-1}$, $\rm {PA_{end\_change}} \geq \rm 1.5PA_{var}$, $R_{b}^0 >$ 2 pixels, $e_{\rm max}$ at the highest peak of $e$ profile).

Using these criteria, we compute an index to identify the presence of a bar, defining a bar as present if the index satisfies more than 6 out of these criteria. These criteria are iteratively tested, manually verified, and refined using hundreds of randomly selected galaxies, ensuring that most cases are accurately addressed. These supplementary criteria help refine the identification process, ensuring that bars are accurately distinguished from other structural features.

We present examples of three types of analysis in Figure \ref{fig:example}. From top to bottom, the panels illustrate a strongly barred galaxy, a weakly barred galaxy, and an unbarred galaxy, respectively. The first two profiles exhibit characteristic bar signatures that satisfy the criteria outlined above, while the third profile does not meet any of these criteria.

\begin{figure*}[!htbp]
\centering
\includegraphics[width=1\linewidth]{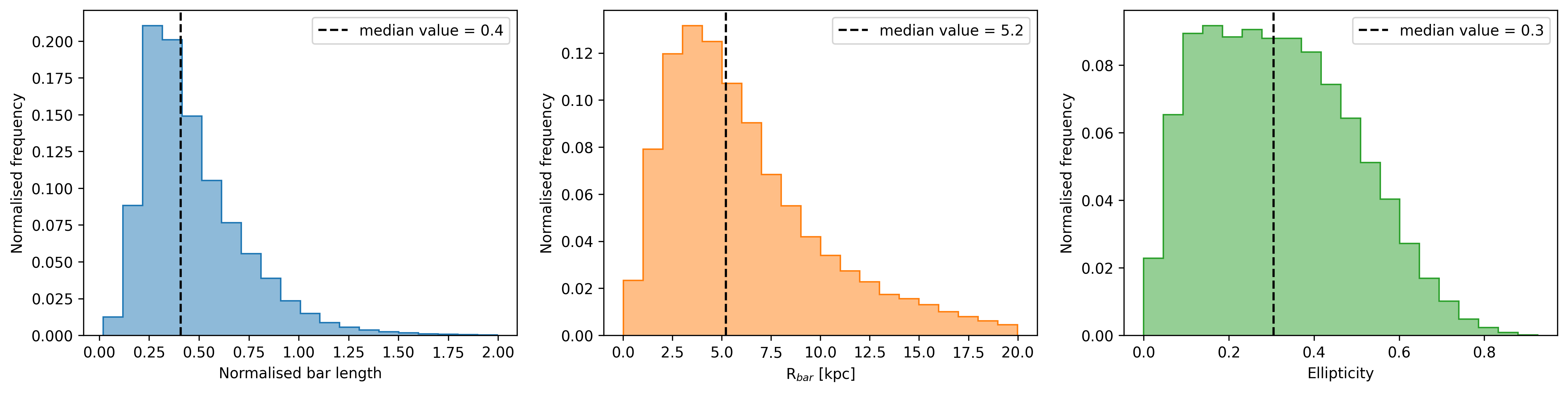}
\caption{Histograms of bar properties. The three histograms display the normalized frequencies of normalized bar length, absolute bar length ($R_{\rm bar}$), and bar ellipticity. The histograms show that most normalized bar lengths fall within the range of 0.25 to 0.6 and most bar ellipticitys fall within the range of 0.2 to 0.5. The black dashed lines indicate the median value of the three parameters: 0.4 for normalized bar length, 5.2 for absolute bar length, 0.3 for bar ellipticity.
\label{fig:hist}}
\end{figure*}

\begin{figure*}[htp!]
\centering
\includegraphics[width=1.0\linewidth]{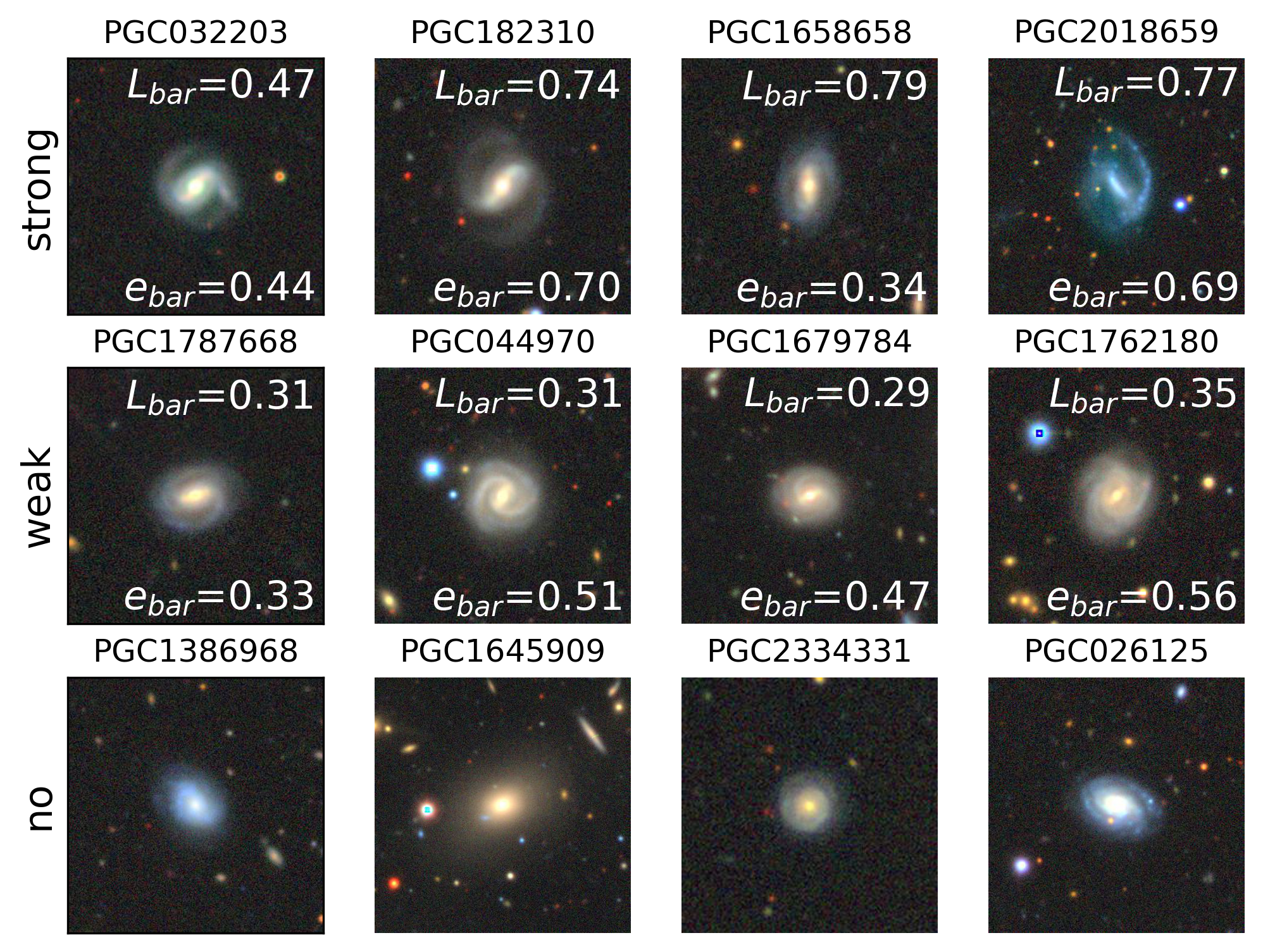}
\caption{DESI postage stamps of 12 example galaxies classified in our study. The normalized bar length ($L_{\rm bar}$) used as the classification criterion for bar strength and bar ellipticity ($e_{\rm bar}$), is shown in the top right and bottom right of each subplot. Galaxies are arranged from top to bottom: strong barred (top four), weak barred (middle four), and unbarred (bottom four).
\label{fig:example2}}
\end{figure*}

\subsection{Measurements}
To investigate bar properties and their effects on galaxies, we calculate the deprojected bar length based on the results obtained from the IRAF ellipse task. When a bar is successfully identified, the semi-major axis corresponding to the peak ellipticity (\textit{e}$_{\rm max}$) is designated as the observed bar length, $R_{b}^0$. To account for the underestimation caused by line-of-sight compression and the overestimation due to spiral arms, we derive the deprojected bar length using the method outlined in \citet{2011ApJS..197...22L}. The deprojected bar length, $R_{b}$, is calculated as:
\begin{equation}\label{eqn-1} 
R_{b}=R_{b}^0\sqrt{(\cos{\Delta PA})^2+(\frac{\sin{\Delta PA}}{(b/a)_g})^2}
\end{equation}
where $\Delta \text{PA}=\text{PA}{\text{g}}-\text{PA}{\text{b}}$, and $\text{PA}{\text{g}}$ and $(b/a){\text{g}}$ represent the position angle and axis ratio of the outer disk of the galaxy, respectively, while $\text{PA}_{\text{b}}$ is the observed position angle of the bar. Similarly, the deprojected bar axis ratio $(b/a)_b^0$ and ellipticity \textit{e}$_{b}$ are also determined using the following formula \citep{2023ApJ...949...91Z}:
\begin{equation}\label{eqn-2} 
(\frac ba)_b=(\frac ba)_{b}^0\sqrt{\frac{((b/a)_g)^2(\cos{\Delta PA_b})^2+(\sin{\Delta PA_b})^2}{((b/a)_g)^2(\cos{\Delta PA_a})^2+(\sin{\Delta PA_a})^2}},
\end{equation}
\begin{equation}\label{eqn-3} 
\textit{e}_{b}=1-(\frac ba)_{b}
\end{equation}
where  the misalignment angle $\Delta\text{PA}_{b}$ and $\Delta\text{PA}_{a}$ measure the difference of position angles between the bar and the major and minor axes of the galaxy disk, respectively. 

Additionally, we calculate the bar length normalized by the disk size of the galaxy, where the disk size is represented by half of the approximate major-axis diameter at the 25 $\rm mag/arcsec^{2}$ optical surface brightness isophote ($R_{25}$). This normalization facilitates a comparative study of bars across galaxies with varying sizes and morphological types.

\subsection{Bar Identification} \label{subsec:bar identification}
We perform isophotal analysis on the $\it g$-, $\it r$-, and $\it z$-band images for each galaxy. From the results obtained using the ellipse fitting procedure, we identify isophotes for 212,555 galaxies in the g band, 212,708 galaxies in the r band, and 212,264 galaxies in the z band. The discrepancies arise primarily due to the lack of imaging data in certain bands within the DESI surveys. Based on the analysis, we determine a total bar fraction of 44.1\% in the $\it r$ band, with similar fractions in the $\it g$ (45.0\%) and $\it z$ (42.4\%) bands.

To ensure the robustness of our identification, we require that bars are detected in at least two bands, reducing the sample size to 210,917 galaxies. Under this criterion, we derive a total bar fraction of 42.9\%. This value is lower than those reported in earlier studies, which suggested that stellar bars exist in approximately two-thirds of nearby disk galaxies. The discrepancy arises because our sample includes not only disk galaxies but also a substantial number of elliptical galaxies.

To focus specifically on disk galaxies, we utilize the GZD catalog and apply the criteria proposed by \citet{2021MNRAS.507.4389G}: $p_{\rm{featured-or-disk}} \geq 0.27$ and $p_{\rm{edge-on\_no}} \geq 0.68$. Here, $p_{\rm{featured-or-disk}}$ represents the fraction of participants who classified the galaxy as having “features or a disc,” indicating the presence of distinguishable morphological structures. Similarly, $p_{\rm{edge-on\_no}}$ refers to the fraction of participants who classified the galaxy as not being edge-on, reflecting its orientation. These criteria help isolate a robust sample of disk galaxies for further analysis.

Using these standards, we identify 102,640 disk galaxies from a cross match sample of 202,322 galaxies, for which the total bar fraction is found to be 62.0\%. This result aligns closely with the widely accepted conclusion regarding bar fractions in disk galaxies.

\subsection{Bar Properties}
The normalized bar length, absolute bar length, and bar ellipticity are statistically summarized in Figure \ref{fig:hist}. As shown in the histograms, most bars are found to have absolute lengths of 3–7 kpc. The normalized bar length is primarily concentrated in the range of 0.25–0.6, while ellipticity is predominantly distributed between 0.2 and 0.5, reflecting the typical range observed in barred galaxies.

The normalized bar length is scaled to $R_{25}$ of the host galaxy. Its median value, marked by a dashed line in the graph, is approximately 0.4. In this work, we use this value as a threshold to classify bars into strong and weak categories: bars with a normalized length greater than 0.4 are categorized as strong, while those with a normalized length less than 0.4 are considered weak. Using this classification, we find that the strong bar fraction is 22.4\% for the entire sample and 30.0\% for the disk galaxy subset.

We present classification results for 12 examples of robustly identified bars in Figure \ref{fig:example2}. From the top row to the bottom row, examples of strong, weak, and non-barred galaxies are displayed. The normalized bar length is indicated in the upper right corner of each panel. For further study, we provide a sample of 20 galaxies in Table \ref{tab:d}, listing bar properties (normalized bar length and ellipticity) and basic galaxy information from the DESI catalog. The complete dataset will be made publicly available online:\dataset[10.57760/sciencedb.19973]{\doi{10.57760/sciencedb.19973}}.

\section{Results} \label{sec:results}

\begin{figure*}[htp!]
\centering
\includegraphics[width=1\linewidth]{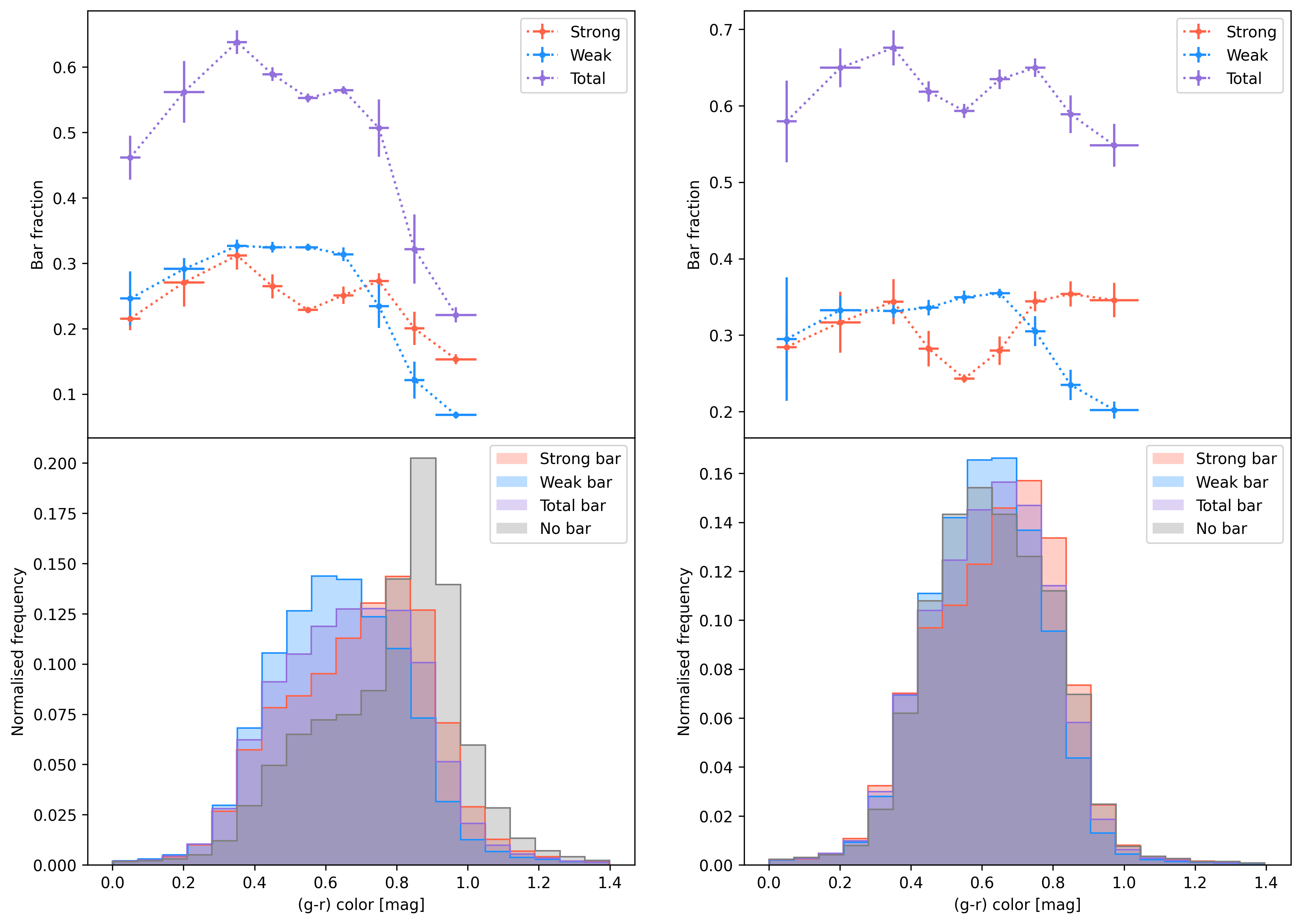}
\caption{Effect of ($\it g - \it r$) Color on Bar Fractions and Distribution.
Top Panels: The influence of ($\it g - \it r$) color on strong (red), weak (blue), and total (purple) bar fractions is shown for the total sample (left) and disk sample (right). Galaxies are grouped into equal-sized color bins, with smaller sub-bins used to calculate the standard deviations of color (horizontal bars) and bar fraction (vertical bars). Bottom Panels: Histograms display the normalized frequencies of strong barred (red), weak barred (blue), total barred (purple), and unbarred (gray) galaxies in the total sample (left) and disk sample (right).
\label{fig:histcolor_both}}
\end{figure*}

\subsection{Distribution of Bar Fraction}
\begin{figure*}[htp!]
\centering
\includegraphics[width=1\linewidth]{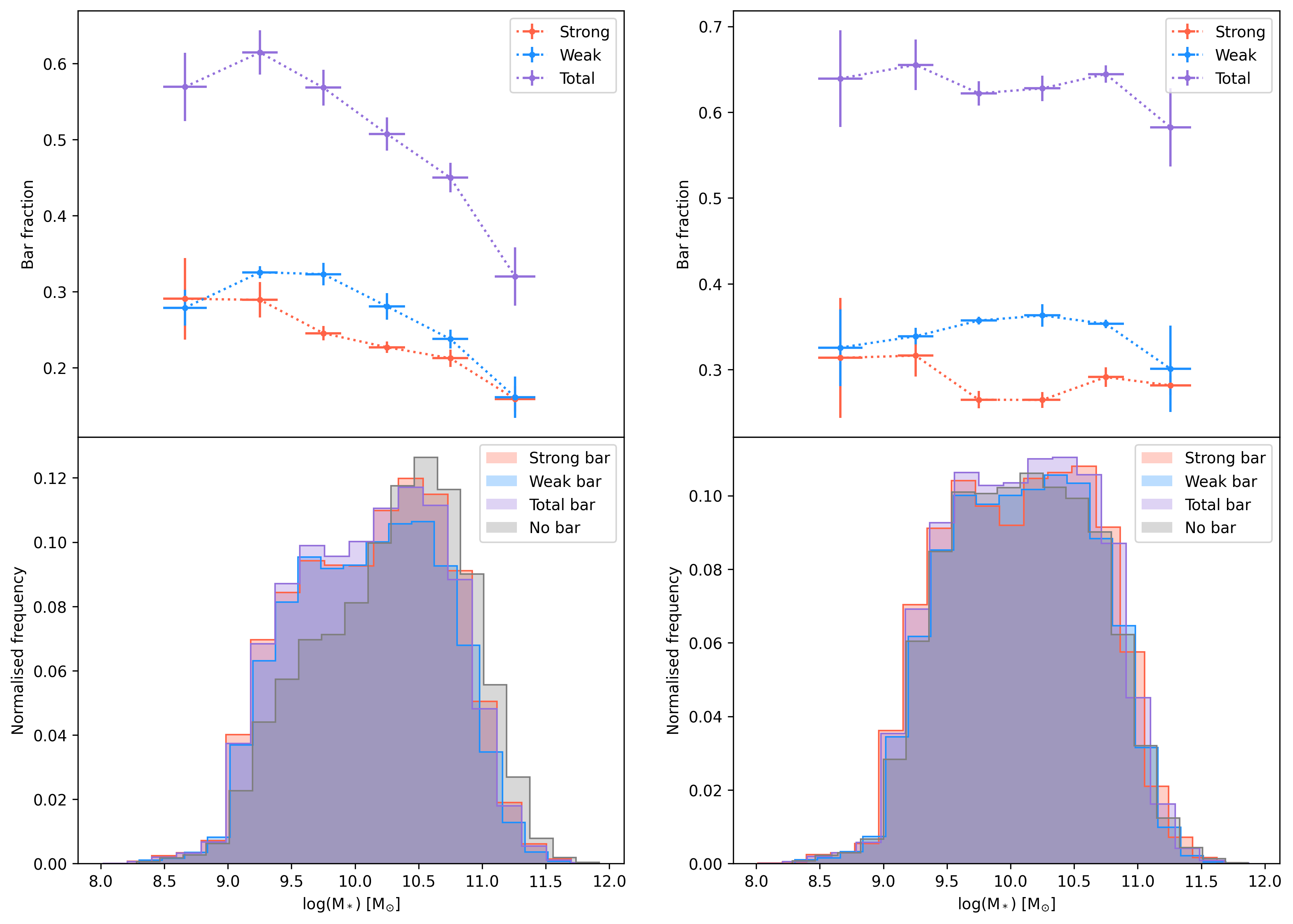}
\caption{Dependence of bar fractions and distribution on stellar mass of galaxies ($0.01 < z < 0.05$ and $M_{r} < -18.96$). Top Panels: The relationship between stellar mass and strong (red), weak (blue), and total (purple) bar fractions is shown for the total sample (left) and disk sample (right). Galaxies are grouped into equal-sized mass bins. Smaller sub-bins within each bin are used to calculate the standard deviations of stellar mass (horizontal bars) and bar fraction (vertical bars). Bottom Panels: Histograms display the normalized frequencies of strong barred (red), weak barred (blue), total barred (purple), and unbarred (gray) galaxies in the total sample (left) and disk sample (right).
\label{fig:histmass_both}}
\end{figure*}

In this section, we analyze the overall distribution of bars in relation to the properties of the host galaxies. The ($\it g - \it r$) color, corrected for Milky Way extinction, is used to represent the overall color of galaxies. Galaxies are categorized into strong or weak bars based on their normalized bar length, and their counts and fractions are computed accordingly. We investigate the bar fractions and distributions across the entire sample and the disk galaxy subset. The results are presented in Figure \ref{fig:histcolor_both} and Figure \ref{fig:histmass_both}. To minimize the bias introduced by the estimation of redshift and stellar mass, we applied volume-limited criteria to the analysis by imposing redshift thresholds ($0.01 < z < 0.05$) and r-band absolute magnitude threshold ($M_{r} < -18.96$) when examining stellar mass in the follow-up analysis. This leads to the sample size of 70,515 for the total sample and 45,581 for the disk sample.

\begin{figure}[htp!]
\centering
\includegraphics[width=1\linewidth]{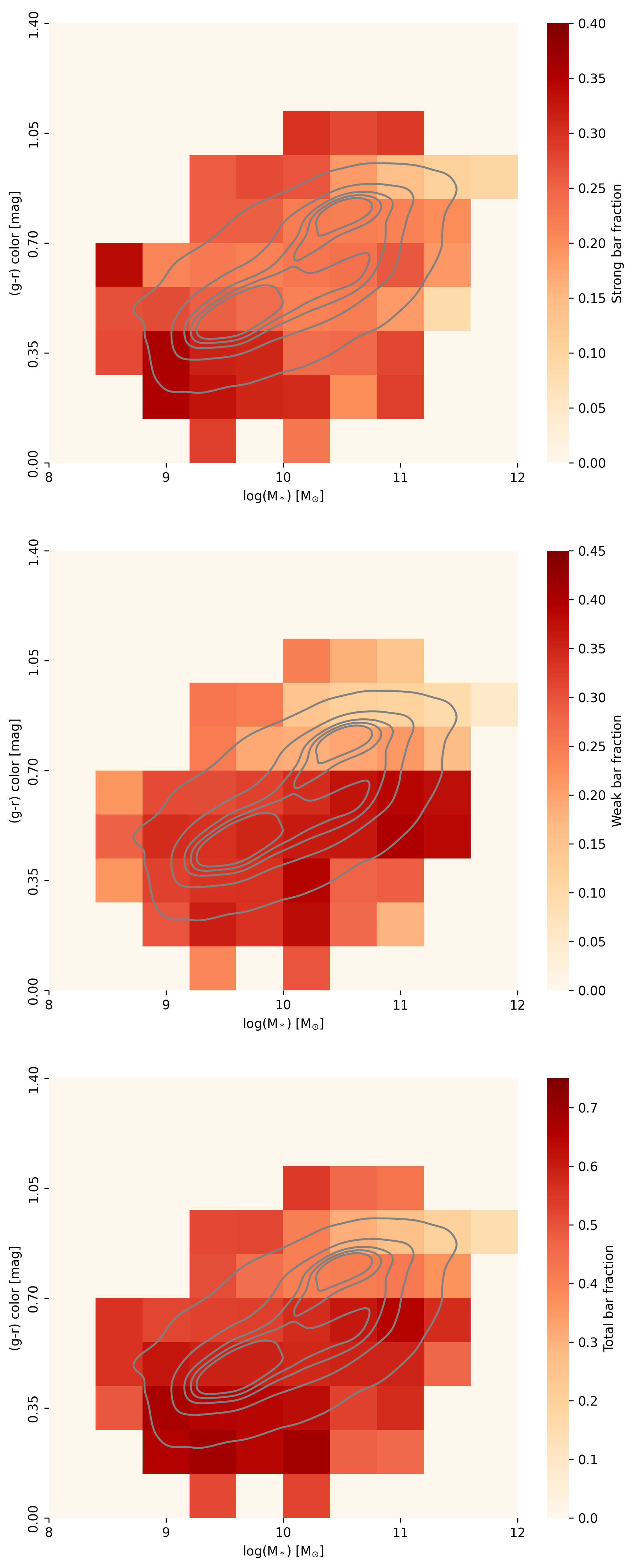}
\caption{Fraction of all barred galaxies in the stellar mass - galaxy ($\it g - \it r$) color plane ($0.01 < z < 0.05$ and $M_{r} < -18.96$). The stellar mass–($\it g - \it r$) color plane is binned along both axes, with the color scale indicating the mean bar fraction within each bin for strong (top panel), weak (middle panel), and total (bottom panel) bars in the total sample. Background contours (gray solid lines) represent percentiles containing galaxies in the total sample. The color intensity reflects the bar fraction value within each bin.
\label{fig:SWT}}
\end{figure}

\begin{figure}[htp!]
\centering
\includegraphics[width=1\linewidth]{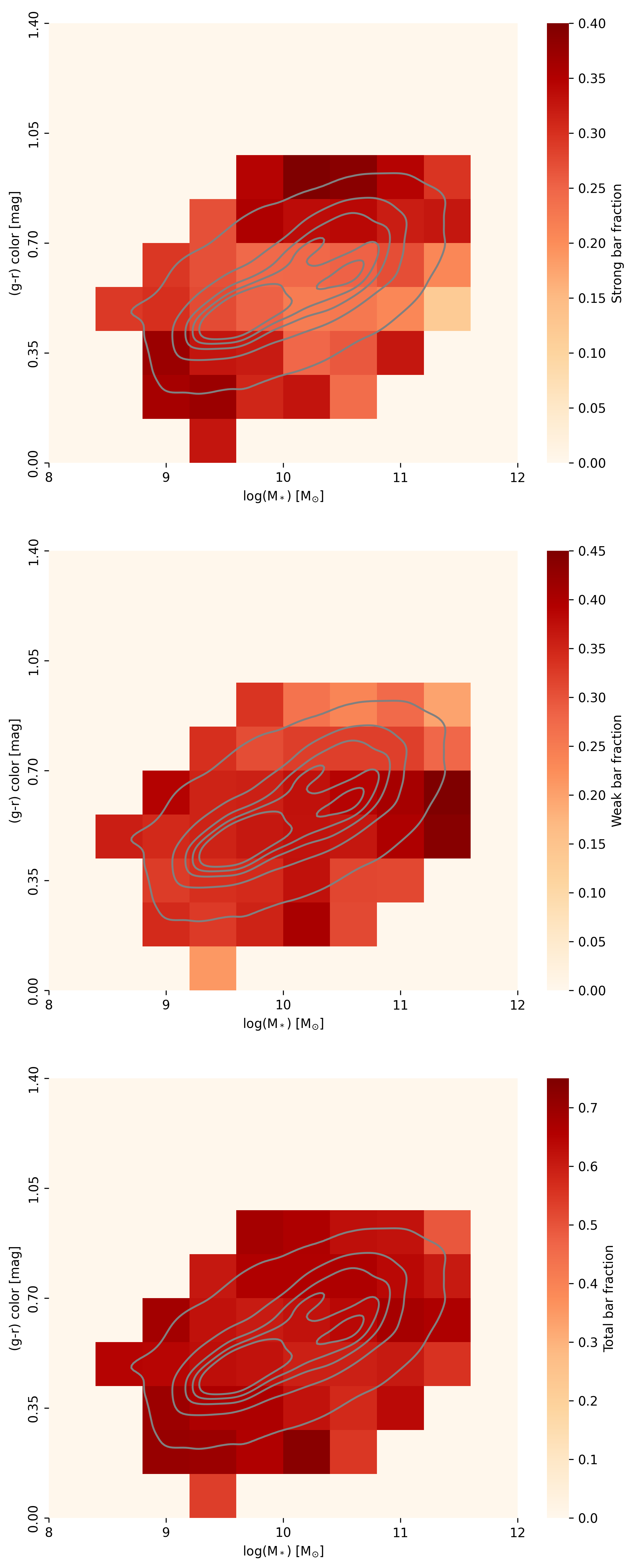}
\caption{Similar to Figure \ref{fig:SWT}, but for the disk galaxy subset.
\label{fig:SWT_disk}}
\end{figure}

Figure \ref{fig:histcolor_both} illustrates the relationship between bar fraction (upper panels) and count frequency (lower panels) as a function of galaxy color. The left plot highlights the distribution of the ($g-r$) color across the entire sample. A distinct bimodal pattern emerges, with total bar fractions peaking above 60\% at $g-r \rm \approx 0.3$ and around 55\% at $g-r \rm \approx 0.6$. Bars are less prevalent in the reddest regions with $g-r \rm \sim 1.0$, largely due to the dominance of massive elliptical galaxies. Strong bars also display a bimodal distribution, with peaks at $g-r \rm \approx 0.3$ and $0.7$. In contrast, weak bar fractions rise steadily up to $g-r \rm = 0.4$ before declining. Barred galaxies exhibit a more uniform distribution across colors, while unbarred galaxies predominantly cluster in the redder range ($g-r \rm \approx 0.8-1.0$). Notably, strong bars tend to be associated with redder galaxies compared to weak bars.

The right panel of Figure \ref{fig:histcolor_both} presents the bar distribution in the disk galaxy subset selected in Section \ref{subsec:bar identification}. Similar to the total sample, the disk galaxies exhibit a bimodal bar fraction distribution, with peaks around $g-r \approx 0.3$ and $0.7$, both exceeding 60\%. Both fractions of strong and weak bars reach higher peaks, compared to the total sample. In the histogram, weak bars are slightly bluer than strong bars, while their unbarred counterparts are more uniformly distributed in the color range after the non-disk galaxies are removed from the total sample.

Figure \ref{fig:histmass_both} (left panel) shows the distribution of total stellar mass across different bar classifications within $0.01 < z < 0.05$ and $M_{r} < -18.96$. In the total sample, bar fractions decline with increasing stellar mass, from around 60\% in low-mass ($M_* < 10^{9}M_{\odot}$) galaxies to 30\% in high-mass ($M_* > 10^{11}M_{\odot}$) galaxies. Both strong and weak bar fractions have similar distributions, while strong bars are associated with more massive galaxies compared to weak bars. The disk galaxy subset in the right panel of Figure \ref{fig:histmass_both} shows weak bimodal distribution for the total bar fraction along with stellar mass. The strong and weak bar fractions also show shallower slopes in terms of the stellar mass. 

\begin{figure*}[htp!]
\centering
\includegraphics[width=1\linewidth]{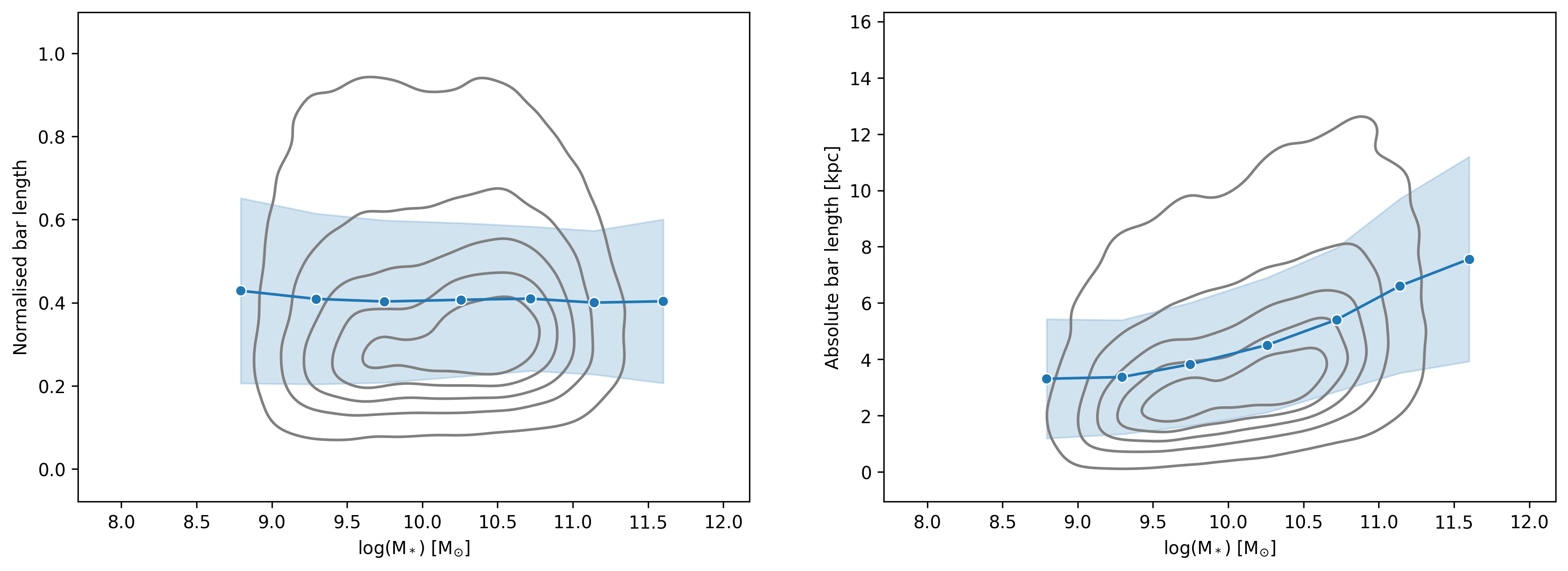}
\caption{Bar length as a function of stellar mass. The normalized bar length (left) and absolute bar length (right) are plotted as functions of stellar mass for barred galaxies ($0.01 < z < 0.05$ and $M_{r} < -18.96$). Background contours (gray solid lines) represent percentiles containing the galaxies. Solid points indicate the mean bar length within each mass bin, with shaded regions representing the standard deviation.
\label{fig:barmass}}
\end{figure*}

To further investigate how bars depend on galaxy properties, we examine the distribution of different bar types on the stellar mass–color plane for both the entire sample and the disk galaxy subset, as shown in Figures \ref{fig:SWT} and \ref{fig:SWT_disk}, respectively. These figures present the fractions of strong, weak, and total bars in the top, middle, and bottom panels, respectively. Each bin spans 0.4 dex in log $M_*$ and 0.14 in $(g-r)$ color and is color-coded to indicate the bar fraction within that bin. To ensure statistical reliability, we only include bins containing at least 50 galaxies.

Figure \ref{fig:SWT} shows that bars are more common in the bluer regions ($g-r < 0.7$) and significantly rarer in the massive, red region ($M_* > 10^{10}M_{\odot}$, $g-r > 0.7$), which is predominantly occupied by quenched elliptical galaxies. To examine the bar distribution without the influence of elliptical galaxies, we focus on the disk galaxy subset.

Figure \ref{fig:SWT_disk} highlights the bimodal distribution of bar fractions in disk galaxies, particularly for strong bars. One peak is observed in the blue, low-mass region ($M_* < 10^{10}M_{\odot}$, $g-r < 0.4$), while the other is in the red, intermediate-mass region ($10^{10}M_{\odot} < M_* < 10^{11}M_{\odot}$, $g-r > 0.7$). Additionally, weak bars are more uniformly distributed across the stellar mass–color plane.

\subsection{Bar Size and Stellar Mass}
We examine the relationship between normalized and absolute bar lengths and stellar mass for galaxies with $0.01 < z < 0.05$ and $M_{r} < -18.96$, as shown in Figure \ref{fig:barmass}. The contours represent the number density of barred galaxies, with bar lengths plotted against stellar mass.

In the left panel, normalized bar length shows minimal variation with stellar mass, generally remaining centered around 0.4. In the right panel, absolute bar length shows a clear positive correlation with stellar mass. Bars grow longer with increasing stellar mass. Specifically, low-mass galaxies ($M_{*} < 10^{10} M_{\odot}$) typically have bars measuring 2-4 kpc, and high-mass galaxies ($M_{*} > 10^{10} M_{\odot}$) can have bars extending to 8 kpc. 

Additionally, the slope of the bar length increase with stellar mass is steeper at higher mass, indicating that in high-mass galaxies, absolute bar length varies more significantly with mass compared to low-mass galaxies. This suggests that bar growth is more strongly influenced by stellar mass in massive galaxies.

\section{Validation \& Discussion} \label{sec:discussion}

\subsection{Identification Inspection}
To validate the accuracy of our bar identification method, we compare our results with deep learning models in GZD. As described in Section \ref{sec:2.2.1}, we utilize the GZD catalog for comparison. It contains predicted posteriors generated by trained neural networks.

\begin{figure}[htp!]
\centering
\includegraphics[width=1\linewidth]{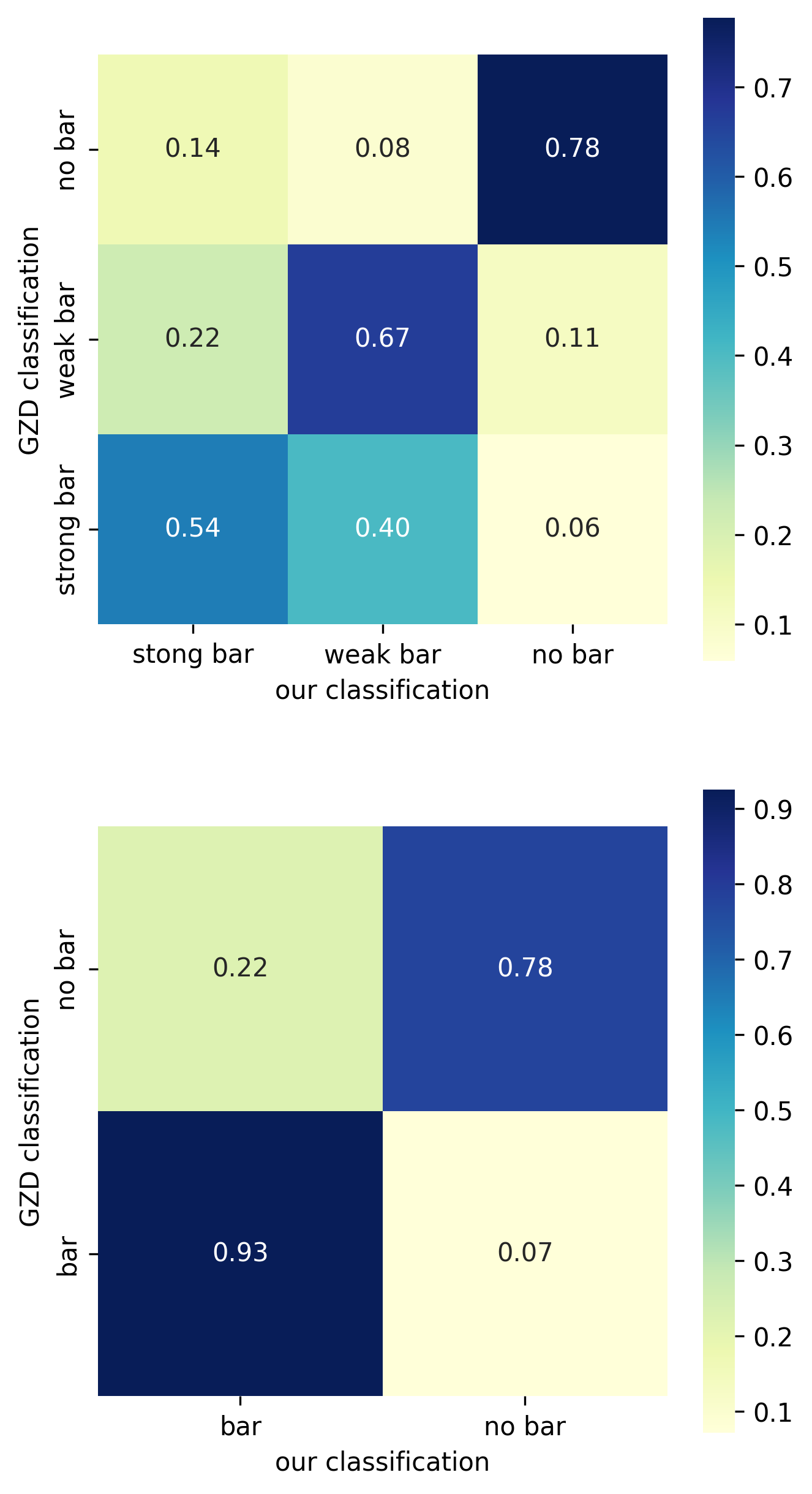}
\caption{Confusion matrices for bar classifications comparing GZD classifications (columns) with our results (rows). Each entry ($\it i$, $\it j$) indicates the fraction of galaxies classified as type $\it i$ in our study that are classified as type $\it j$ in GZD. The top panel compares strong, weak, and no bars, while the bottom panel combines strong and weak bars to assess total bar identification accuracy.
\label{fig:GDZ-auto}}
\end{figure}

For consistency, we apply a fraction threshold of 0.8 to classify bars with higher confidence, Strong bars: $p_{\rm{bar\_strong}} > p_{\rm{bar\_weak}}$ and $1 - p_{\rm{not\_bar}} \geq 0.8$, Weak bars: $p_{\rm{bar\_weak}} > p_{\rm{bar\_strong}}$ and $1 - p_{\rm{not\_bar}} \geq 0.8$, Non-bars: $p_{\rm{not\_bar}} \geq 0.8$. To select total bars (including both strong and weak bars), we use the threshold of $1 - p_{\rm{not\_bar}} \geq 0.8$, where $p_{\rm{bar\_strong}} + p_{\rm{bar\_weak}} = 1 - p_{\rm{not\_bar}}$. The confusion matrices comparing our classifications with those in GZD are shown in Figure \ref{fig:GDZ-auto}.

GZD demonstrates a strong consistency with our bar classifications. As shown in the confusion matrix (top panel of Figure \ref{fig:GDZ-auto}), we achieve verification rates of 54\% for strong bars, 67\% for weak bars, and 78\% for non-bars. Furthermore, 94\% of the strong bars and 89\% of the weak bars identified by GZD are correctly classified as either strong or weak bars in our results. When combining strong and weak bars to evaluate the total bar fraction (bottom panel of Figure \ref{fig:GDZ-auto}), the overall accuracy improves to 93\%, highlighting the robustness and reliability of our bar identification methodology.

A notable inconsistency lies in our classification of strong bars compared to GZD. This discrepancy may stem from differences between our classification standards and those used in visual inspection. This observation suggests that classifying bar strength is inherently challenging. One key difficulty is the absence of a clear and widely accepted definition distinguishing strong and weak bars. This uncertainty affects both visual and quantitative methods.


\begin{figure*}[htp!]
\centering
\includegraphics[width=1\linewidth]{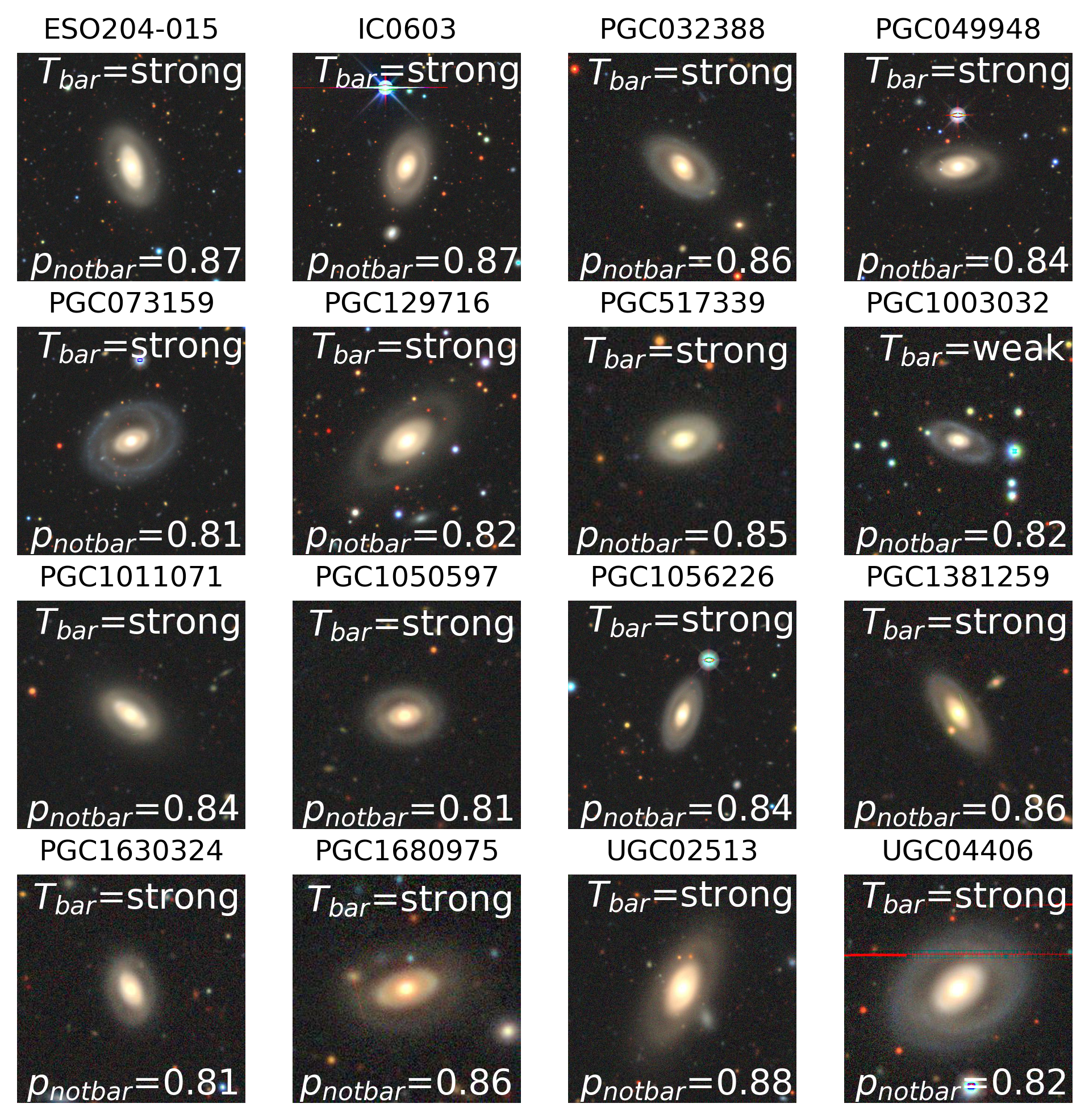}
\caption{DESI postage stamps of 16 galaxies classified as unbarred by GZD but identified as barred in our study. The classification in our study is shown in the upper right of each panel, while the probability of being unbarred ($p_{\rm notbar}$) from GZD is displayed at the bottom of each panel.
\label{fig:nb}}
\end{figure*}

Additionally, we find that approximately 22\% of galaxies classified as unbarred in GZD are identified as barred in our results. This discrepancy highlights the necessity of combining qualitative and quantitative methods, which can overcome the limitations of human visual inspection and improve classification accuracy. To address this, we further examine galaxies that are classified as barred in our study but as unbarred in GZD. Figure \ref{fig:nb} presents randomly selected examples from this group. Most of these galaxies display distinct bar- or oval-like structures and disk features in the DESI imaging data. However, it should be noted that some bar-like structures may be misidentified as bars. This highlights the challenges in distinguishing true bars from similar structures with quantitative methods, which may contribute to the inconsistencies and classification mismatches observed in Figure \ref{fig:GDZ-auto}. Therefore, visual approaches remain essential for providing additional context and improving classification accuracy.

\subsection{Correlations of Bar Length and Strength}
In this study, we use the normalized bar length relative to the host galaxy as a proxy for bar strength, classifying bars as strong or weak based on a threshold of 0.4, corresponding to the median value. Although some discrepancies exist between our classification and those based on subjective visual inspection, bar size remains a reasonable indicator of bar type and strength.

Numerical simulations and observations in the local Universe consistently demonstrate that bar length traces both bar evolution and strength. Theoretical models predict that bars grow longer and stronger as they evolve through angular momentum transfer from the bar to the outer disk \citep{2010ApJ...719.1470V, 2013ApJ...779..162C, 2013seg..book..305A}. Empirical studies further support this, showing a positive correlation between bar length and strength \citep{2004AJ....128..183B, 2007ApJ...670L..97E, 2007ApJ...657..790M}. Specifically, \citet{2016A&A...587A.160D} found that both absolute and normalized bar lengths correlate with Fourier density amplitude, indicating that longer bars are indeed stronger. Strong bars are generally longer than weak bars, with normalized lengths covering approximately 40\%–45\% of the host galaxy for strong bars and 20\%–30\% for weak bars \citep{2021MNRAS.507.4389G}.

As \citet{2021MNRAS.507.4389G} suggested, distinguishing between weak and strong bars is useful for understanding their different properties and effects on host galaxies, even though bars exist on a continuum rather than as distinct categories. Classifying bars based on length captures these extremes, allowing for meaningful insights into bar-driven secular evolution and its impact on galaxy morphology. Observations also confirm that longer bars are found in larger galaxies, highlighting the link between bar size and host galaxy size \citep{2009A&A...495..491A}.

Thus, setting the classification threshold at a normalized bar length of 0.4 is both practical and consistent with previous studies. This approach effectively captures the trends in bar evolution and strength, aligning with theoretical predictions and observational evidence. However, it is important to note that, different methods, criteria, or conditions may lead to variations in bar classification \citep{2019ApJ...872...97L}. In our study, we classify bars based on normalized bar length. This approach differs from other studies that use purely visual methods. As a result, the strong bars in our analysis likely include both strong and weak bars as defined in those studies.

\subsection{Correlations of Bars and Host Galaxy Properties}
The formation, evolution, and properties of bars are intricately tied to the characteristics of their host galaxies, including color, stellar mass, and redshift. Numerous studies have explored these relationships, offering valuable insights into how bars both influence and reflect galaxy evolution \citep{2003MNRAS.341.1179A, 2018MNRAS.474.5372E, 2022MNRAS.512.5339R}.

Figures \ref{fig:histcolor_both} and \ref{fig:SWT} reveal a clear bimodal distribution of bar fractions with respect to galaxy color, aligning with earlier findings of bimodal distributions in galaxy color and morphology \citep{2008ApJ...675.1194B, 2011MNRAS.411.2026M, 2021MNRAS.507.4389G}. Specifically, strong bars tend to reside in redder galaxies, while weak bars are more common in bluer systems (Figure \ref{fig:histcolor_both}). This trend is consistent with \citet{2012ApJ...745..125L}, who found that weak bars predominantly occur in late-type, low-mass, blue spirals, whereas strong bars peak in redder galaxies.

These patterns highlight the complex interplay between bar strength, stellar population, and galaxy morphology. Observational evidences indicate that barred galaxies are not homogeneous in their stellar populations \citep{2011MNRAS.415..709S, 2016A&A...595A..63V}. Strong bars are generally associated with redder and older stellar populations compared to weak bars \citep{2020MNRAS.495.4158F, 2023A&A...671A...8D}. Bars in bluer galaxies may reflect early evolutionary stages, while bars in redder galaxies are more evolved \citep{2011MNRAS.415.3627H, 2012MNRAS.423.3486W}.

Multiple studies have shown that bars grow longer and stronger as galaxies evolve \citep{2013MNRAS.432L..56S, 2016A&A...587A.160D}. As illustrated in Figure \ref{fig:barmass}, normalized bar length remains relatively stable across different stellar masses, suggesting that bar size scales proportionally with galaxy size. In contrast, absolute bar length increases with stellar mass at fixed redshifts, indicating that larger galaxies host longer bars, particularly at higher redshifts. Previous studies have also confirmed a positive correlation between absolute bar length and galaxy mass \citep{2016A&A...587A.160D}. Observational studies extending to high redshifts have found no significant changes in normalized bar length with redshift or stellar mass, further supporting the idea that bars scale proportionally with their host galaxies as they evolve over time \citep{2021ApJ...922..196K, 2024MNRAS.530.1171C}.

These results support the theory of secular evolution, where bars grow through angular momentum transfer from the bar to the outer disk, becoming stronger and longer over time \citep{2003MNRAS.341.1179A}. This process not only impacts the bar itself but also drives changes in the host galaxy’s structure and dynamics.

As mentioned above, bars also play a critical role in shaping galaxy morphology and dynamics. In more evolved systems, longer bars act as efficient channels for gas inflow, potentially enhancing or quenching central star formation and driving structural transformations \citep{2010ApJ...719.1470V, 2020MNRAS.495.4158F}. However, certain conditions can suppress bar growth or even destroy existing bars. For example, the development of large bulges or depleted gas reservoirs may inhibit bar growth \citep{2008ASPC..396..351B,2018MNRAS.475.1653K}. Conversely, in low-mass, gas-rich galaxies, excessive gas content can weaken or delay bar evolution \citep{2013MNRAS.429.1949A, 2018MNRAS.474.5372E}.

These findings underscore the complexity of bar distribution and characteristics, influenced by multiple factors such as host galaxy properties, sample selection, and analysis methods \citep{2024MNRAS.530.1171C}.  Addressing these challenges requires larger, more diverse datasets. Upcoming wide-field surveys, such as LSST \citep{2019ApJ...873..111I}, Euclid \citep{2022A&A...662A.112E}, Roman \citep{2019arXiv190205569A}, and CSST \citep{2011SSPMA..41.1441Z, 2019ApJ...883..203G}, will provide deeper insights into bar formation and evolution, helping to clarify these complex relationships.

\section{Summary \& Conclusions} \label{sec:summary}
In this study, we applied an improved isophote analysis to identify bars in 232,142 galaxies imaged by the DESI Legacy Imaging Surveys, discovering nearly 100,000 barred galaxies in the local Universe. We then analyzed the physical properties of the bars, including their fractions, sizes, ellipticities, and correlations with host galaxy characteristics. Our main results are summarized as follows:

(1) Based on the isophote fitting on DESI $\it g$, $\it r$, $\it z$ images, we identified bars and measured their deprojected absolute lengths and ellipticities. The overall bar fraction in our sample is 42.9\%. To classify bars, we normalized bar lengths by galaxy sizes and set a threshold to distinguish strong and weak bars. We use the median values of 0.4. Based on this classification, we found that 22.3\% of galaxies host strong bars. Focusing on disk galaxies, which we selected to ensure more reliable results, we obtained a total bar fraction of 62.0\% and a strong bar fraction of 30.0\%.

(2) We compared our results with visual classifications from the GZD catalog. For bar identification, we achieved an accuracy of 93\% compared to GZD. The classification accuracy for strong and weak bars was 54\% and 67\% in GZD. These discrepancies can be attributed to the differences of identification methodology and the inherent challenges of distinguishing between strong and weak bars.

(3) Bar lengths are measured both in absolute terms and normalized by galaxy size. Most bars are found to have absolute lengths of 3–7 kpc, and normalized bar lengths concentrated around a median value of 0.4. Bar ellipticity primarily ranges from 0.2 to 0.6, with a median value of 0.3.

(4) We observed a bimodal distribution of bar fractions with respect to galaxy color, peaking at $g-r \sim 0.3$ and $g-r \sim 0.7$ in disk galaxies. Bar fractions have remained consistently high, showing slight increases at specific colors. Strong bars show a similar bimodal trend as well, while weak bars peak at bluer colors and decrease in redder galaxies. Unbarred galaxies are concentrated in reddest regions ($g-r \sim 1.0$) in our sample.

(5) In the overall sample, bar fractions decrease with increasing stellar mass due to the presence of elliptical galaxies. In the disk galaxy subset, the total bar fraction exhibits a weak bimodal distribution with stellar mass, with peaks at $M_{*} < 10^{9} M_{\odot}$ and $M_{*} \sim 10^{11} M_{\odot}$.

(6) On the stellar mass–color plane, bars in disk galaxies, particularly strong bars, are more common in intermediate-mass, red galaxies and low-mass, blue galaxies, while weak bars are uniformly distributed, likely reflecting differences in their formation and evolutionary pathways.

(7) The normalised bar length remains almost unchanged with larger stellar masses, indicating that bar size scales proportionally with galaxy size. Absolute bar length correlated positively with stellar mass of host galaxies, with longer bars found in more massive galaxies.

\section*{Acknowledgements}
\label{sec:acknow}
We sincerely thank the anonymous referee and the editor for their critical comments and instructive suggestions, which have significantly enhanced the quality of this paper. This work was supported by the National Key Research and Development Program of China (No. 2022YFA1602902), by the Chinese National Natural Science Foundation grant (Nos. 12073035).

The DESI Legacy Imaging Surveys consist of three individual and complementary projects: the Dark Energy Camera Legacy Survey (DECaLS), the Beijing-Arizona Sky Survey (BASS), and the Mayall z-band Legacy Survey (MzLS). DECaLS, BASS and MzLS together include data obtained, respectively, at the Blanco telescope, Cerro Tololo Inter-American Observatory, NSF’s NOIRLab; the Bok telescope, Steward Observatory, University of Arizona; and the Mayall telescope, Kitt Peak National Observatory, NOIRLab. NOIRLab is operated by the Association of Universities for Research in Astronomy (AURA) under a cooperative agreement with the National Science Foundation. Pipeline processing and analyses of the data were supported by NOIRLab and the Lawrence Berkeley National Laboratory (LBNL). Legacy Surveys also uses data products from the Near-Earth Object Wide-field Infrared Survey Explorer (NEOWISE), a project of the Jet Propulsion Laboratory/California Institute of Technology, funded by the National Aeronautics and Space Administration. Legacy Surveys was supported by: the Director, Office of Science, Office of High Energy Physics of the U.S. Department of Energy; the National Energy Research Scientific Computing Center, a DOE Office of Science User Facility; the U.S. National Science Foundation, Division of Astronomical Sciences; the National Astronomical Observatories of China, the Chinese Academy of Sciences and the Chinese National Natural Science Foundation. LBNL is managed by the Regents of the University of California under contract to the U.S. Department of Energy. The complete acknowledgments can be found at https://www.legacysurvey.org/acknowledgment/.


\bibliography{sample631}{}
\bibliographystyle{aasjournal}



\end{document}